\documentclass[preprint,12pt]{elsarticle}

\usepackage{amssymb}

\newcommand\rf[1]{(\ref{eq:#1})}
\newcommand\lab[1]{\label{eq:#1}}
\newcommand\nonu{\nonumber}
\newcommand\br{\begin{eqnarray}}
\newcommand\er{\end{eqnarray}}
\newcommand\be{\begin{equation}}
\newcommand\ee{\end{equation}}

\newcommand\foot[1]{\footnotemark\footnotetext{#1}}
\newcommand\lb{\lbrack}
\newcommand\rb{\rbrack}

\newcommand\llb{\left\lbrack}
\newcommand\rrb{\right\rbrack}

\newcommand\lcurl{\left\{}
\newcommand\rcurl{\right\}}
\renewcommand\({\left(}
\renewcommand\){\right)}
\newcommand\bgv{\bigg\vert}              

\newcommand\bc{\begin{center}}
\newcommand\ec{\end{center}}

\newcommand\partder[2]{\frac{{\partial {#1}}}{{\partial {#2}}}}

\renewcommand\a{\alpha}
\renewcommand\b{\beta}

\renewcommand\d{\delta}

\newcommand\vareps{\varepsilon}
\newcommand\g{\gamma}
\newcommand\G{\Gamma}

\newcommand\h{\frac{1}{2}}
\renewcommand\k{\kappa}
\renewcommand\l{\lambda}
\renewcommand\L{\Lambda}
\newcommand\m{\mu}
\newcommand\n{\nu}

\newcommand\vp{\varphi}

\newcommand\pa{\partial}

\newcommand\pr{\prime}

\renewcommand\th{\theta}

\newcommand\cE{{\mathcal E}}

\newcommand\cM{{\mathcal M}}

\newcommand{\ct}[1]{\cite{#1}}
\newcommand{\bib}[1]{\bibitem{#1}}

\newcommand\PRL[3]{\textsl{Phys. Rev. Lett.} \textbf{#1} (#2) #3}
\newcommand\NPB[3]{\textsl{Nucl. Phys.} \textbf{B#1} (#2) #3}

\newcommand\PRD[3]{\textsl{Phys. Rev.} \textbf{D#1} (#2) #3}

\newcommand\PLB[3]{\textsl{Phys. Lett.} \textbf{#1B} (#2) #3}
\newcommand\CQG[3]{\textsl{Class. Quantum Grav.} \textbf{#1} (#2) #3}

\newcommand\AoP[3]{\textsl{Ann. of Phys.} \textbf{#1} (#2) #3}

\newcommand\IJMPA[3]{\textsl{Int. J. Mod. Phys.} \textbf{A#1} (#2) #3}
\newcommand\IJMPD[3]{\textsl{Int. J. Mod. Phys.} \textbf{D#1} (#2) #3}

\newcommand\MPLA[3]{\textsl{Mod. Phys. Lett.} \textbf{A#1} (#2) #3}

\begin{document}

\begin{frontmatter}



\title{Dynamical Couplings, Dynamical Vacuum Energy and
Confinement/Deconfinement from $R^2$-Gravity}


\author[BGU]{Eduardo Guendelman\corref{cor1}}
\ead{guendel@bgu.ac.il}
\ead[url]{http://eduardo.hostoi.com}
\cortext[cor1]{Corresponding author -- tel. +972-8-647-2508, fax +972-8-647-2904.}
\author[BGU]{Alexander Kaganovich}
\ead{alexk@bgu.ac.il}
\ead[url]{http://profiler.bgu.ac.il/frontoffice/ShowUser.aspx?id=1249}
\address[BGU]{Department of Physics, Ben-Gurion University of the Negev,
P.O.Box 653, IL-84105 ~Beer-Sheva, Israel}

\author[INRNE]{Emil Nissimov}
\ead{nissimov@inrne.bas.bg}
\ead[url]{http://theo.inrne.bas.bg/~nissimov/}
\author[INRNE]{Svetlana Pacheva}
\ead{svetlana@inrne.bas.bg}
\ead[url]{http://theo.inrne.bas.bg/~svetlana/}
\address[INRNE]{Institute for Nuclear Research and Nuclear Energy, Bulgarian Academy
of Sciences, Boul. Tsarigradsko Chausee 72, BG-1784 ~Sofia, Bulgaria}

\begin{abstract}
We study within Palatini formalism an $f(R)$-gravity with $f(R)= R + \a R^2$ 
interacting with a dilaton and a special kind of nonlinear gauge field system 
containing a square-root of the standard Maxwell term, which is known to produce 
confinement in flat space-time. Reformulating the model in
the physical Einstein frame we find scalar field effective potential with a flat
region where the confinement dynamics disappears, while in other regions it
remains intact. The effective gauge couplings as well as the induced
cosmological constant become dynamical. In particular, a conventional
Maxwell kinetic term for the gauge field is
dynamically generated even if absent in the original theory. We find
few interesting classes of explicit solutions: (i) asymptotically (anti-)de Sitter 
black holes of non-standard type with additional confining vacuum electric
potential even for the electrically neutral ones; (ii) non-standard
Reissner-Nordstr{\"om} black holes with additional constant vacuum electric
field and having non-flat-spacetime ``hedgehog'' asymptotics; (iii) generalized 
Levi-Civitta-Bertotti-Robinson ``tube-like'' space-times.
\end{abstract}

\begin{keyword}
Palatini formalism \sep $f(R)$-gravity \sep confining nonlinear gauge field system
\sep dynamical couplings \sep dynamical cosmological constant \sep black
holes with confining electric potential \sep generalized
Levi-Civitta-Bertotti-Robinson spacetimes
\PACS 11.25.-w \sep 04.70.-s \sep 04.50.+h
\end{keyword}

\end{frontmatter}


\section{Introduction}
\label{intro}

In his analysis in refs.\ct{tHooft} `t Hooft has shown that in any effective 
quantum gauge theory, which is able to describe linear confinement phenomena, 
the energy density of electrostatic field configurations should be a linear function
of the electric displacement field in the infrared region. In particular, `t Hooft
has developed a consistent quantum approach in which the electric displacement field
appears as an ``infrared counterterm'' (see especially Eq.(5.10) in second 
ref.\ct{tHooft}).

The simplest way to realize these ideas in flat space-time was proposed in
refs.\ct{GG} by considering the following special nonlinear gauge theory:
\br
S = \int d^4 x L(F^2) \quad ,\quad
L(F^2) = -\frac{1}{4} F^2 - \frac{f_0}{2} \sqrt{-F^2} \; ,
\lab{GG-flat} \\
F^2 \equiv F_{\m\n} F^{\m\n} \quad ,\quad 
F_{\m\n} = \pa_\m A_\n - \pa_\n A_\m  \; ,
\nonu
\er
It has been shown in the first three refs.\ct{GG} that the square root of the Maxwell 
term naturally arises as a result of spontaneous breakdown of scale symmetry of 
the original scale-invariant Maxwell action with $f_0$ appearing as an integration 
constant responsible for the latter spontaneous breakdown. For static field
configurations the model \rf{GG-flat} yields an electric displacement field
$\vec{D} = \vec{E} - \frac{f_0}{\sqrt{2}}\frac{\vec{E}}{|\vec{E}|}$ and 
the corresponding  energy density turns out to be 
$\h \vec{E}^2 = \h |\vec{D}|^2 + \frac{f_0}{\sqrt{2}} |\vec{D}|+\frac{1}{4} f_0^2$, 
so that it indeed contains a term linear w.r.t. $|\vec{D}|$.
The model \rf{GG-flat} produces, when coupled to quantized fermions, a confining 
effective potential $V(r) = - \frac{\b}{r} + \g r$ (Coulomb plus linear one with 
$\g \sim f_0$, see first ref.\ct{GG}) which is of the form of the well-known
``Cornell'' potential \ct{cornell-potential} in 
the phenomenological description of quarkonium systems. 

To this end it is
essential to stress that the Lagrangian $L(F^2)$ \rf{GG-flat} contains both the 
usual Maxwell term as well as the non-analytic square-root function of $F^2$
and thus it is a {\em non-standard} form of nonlinear electrodynamics. 
It is significantly different from the original ``square root''
Lagrangian $- \frac{f}{2}\sqrt{F^2}$ first proposed by Nielsen and Olesen \ct{N-O} to
describe string dynamics. Also it is important that the square root term in 
\rf{GG-flat} is in the ``electrically'' dominated form ($\sqrt{-F^2}$) unlike the 
``magnetically'' dominated Nielsen-Olesen form ($\sqrt{F^2}$).

Let us remark that one could start with the non-Abelian version of the action 
\rf{GG-flat}. Since we will be interested in static spherically symmetric solutions, 
the non-Abelian theory effectively reduces to an Abelian one as pointed out in
the first ref.\ct{GG}.

Coupling of the nonlinear gauge field system \rf{GG-flat} to ordinary Einstein gravity 
was recently studied in \ct{grav-cornell-hiding-hide-confine} 
where the following interesting new features of the pertinent static spherically 
symmetric solutions have been found:

(i) Appearance of a constant radial vacuum electric field (in addition to the
Coulomb one) in charged black holes within Reissner-Nordstr{\"o}m-\-(anti-)\-de-Sitter
space-times, in particular, in electrically neutral black holes with 
Schwarzschild-(anti-)de-Sitter geometry. 

(ii) Novel mechanism of {\em dynamical generation} of cosmological constant
through the nonlinear gauge field dynamics of the ``square-root'' gauge field
term: $\L_{\rm eff} = \L_0 + 2\pi f_0^2$ with $\L_0$ being the bare
cosmological constant.

(iii) In case of vanishing effective cosmological constant 
($\L_0 < 0\; ,\; |\L_0|=2\pi f_0^2$) the resulting
Reissner-Nordstr{\"o}m-type black hole, apart from carrying an additional
constant vacuum electric field, turns out to be {\em non-asymptotically flat} -- 
a feature resembling the gravitational effect of a hedgehog \ct{hedgehog}.

(iv) Appearance of confining-type effective potential in charged test particle 
dynamics in the above black hole backgrounds (cf. Eq.\rf{lin-conf} below).

(v) New ``tube-like'' solutions of Levi-Civita-Bertotti-Robinson \ct{LC-BR} type, 
\textsl{i.e.}, with space-time geometry of the form $\cM_2 \times S^2$, 
where $\cM_2$ is a two-dimensional anti-de Sitter, Rindler or de Sitter space 
depending on the relative strength of the electric field w.r.t. the coupling $f_0$
of the square-root gauge field term.

Let us also mention the recent paper \ct{halilsoy} where coupling of
ordinary Einstein gravity to the  pure ``square-root'' gauge field Lagrangian
($L(F^2) = - \frac{f_0}{2} \sqrt{-F^2}$) is discussed. A new interesting
feature in this model is the existence of dyonic solutions. 

In the present note we will consider $f(R)$-gravity\foot{For a recent review of 
$f(R)$-gravity see \textsl{e.g.} \ct{f(R)-grav} and references therein.
The first $R^2$-model (within the second order formalism), which was also the first
inflationary model, was proposed in \ct{starobinski}.}
with the simplest nonlinear $f(R)= R + \a R^2$ coupled to dilaton $\phi$ and the 
``square-root'' nonlinear gauge system \rf{GG-flat}. 
The main purpose is to study possible new features due
to the combined effect of the simultaneous presence of both the
$R^2$-gravity term and the confinement-producing ``square-root'' of the Maxwell term.
This will be achieved by starting within the first-order (Palatini) formulation of
$f(R)$-gravity (\textsl{e.g.} \ct{f(R)-grav,palatini}) and systematically 
deriving the effective action of the full theory in the physical ``Einstein'' frame. 
The resulting Einstein-frame action is of the form of standard Einstein gravity
interacting with a dilaton $\phi$ and the special nonlinear gauge system with the
square-root term \rf{GG-flat}, where all couplings become dynamically dependent 
on $\phi$. The latter include the effective strength of the standard Maxwell term,
the effective coupling constant of the square-root gauge field term and the
effective cosmological constant. This, in particular, means:

(1) If we start with {\em no} standard Maxwell kinetic term in the original
theory, a nontrivial Maxwell Lagrangian term will nevertheless be
dynamically generated with a $\phi$-dependent strength in the ``Einstein'' frame.
The same is true about the dynamically generated cosmological constant 
out of a zero bare one.

(2) For certain regions of values for the constant dilaton $\phi$ the
effective coupling constant $f_{\rm eff}(\phi)$ of the ``square-root'' 
gauge field term will be vanishing indicating confinement/deconfinement
transition (let us recall that the coupling constant $f_0$ in \rf{GG-flat} 
measures the strength of the effective linear confining potential, see first
ref.\ct{GG} and Eq.\rf{lin-conf} below). 
The above mentioned regions also correspond to flat regions of the
effective scalar potential. The latter could be used as regions in field
space where an inflationary phase for the universe took place according to
the requirements in the ``new inflationary'' scenarios \ct{inflate-phase}.

(3) Furthermore, we find generalizations of the black hole and ``tube-like''
space-time solutions mentioned in (v) above where now their parameters are 
$\phi$-dependent.

We particularly stress that both effects (1)-(2) are entirely due to
both parameters $\a$ (of $R^2$-gravity) and $f_0$ (of nonlinear
``square-root'' gauge theory) simultaneously being non-zero.

Let us also mention a recent work \ct{rubiera} where a $R^2$-gravity interacting
with Born-Infeld nonlinear electrodynamics has been studied and new types of
black hole solutions with different structure of the horizons and
singularities have been found.

\section{Derivation of Einstein-Frame Effective Action of
$R^2$-Gravity-Matter System}
\label{einstein-frame}

The action describing the  coupling of $f(R)= R + \a R^2$ gravity (possibly
with a bare cosmological constant $\L_0$) to a dilaton $\phi$ and
the nonlinear gauge field system with a square-root of the Maxwell term 
\rf{GG-flat} known to produce QCD-like confinement in flat space-time \ct{GG} is 
given by:
\br
S = \int d^4 x \sqrt{-g} \Bigl\lb \frac{1}{2\k^2} 
\Bigl( f\bigl(R(g,\G)\bigr) - 2\L_0 \Bigr) + L(F^2(g)) + L_D (\phi,g) \Bigr\rb \; ,
\lab{f-gravity+GG+D} \\
f\bigl(R(g,\G)\bigr) = R(g,\G) + \a R^2(g,\G) \quad ,\quad 
R(g,\G) = R_{\m\n}(\G) g^{\m\n} \; ,
\lab{f-gravity} \\
L(F^2(g)) = - \frac{1}{4e^2} F^2(g) - \frac{f_0}{2} \sqrt{\vareps F^2(g)} \; ,
\lab{GG-g} \\
F^2(g) \equiv F_{\k\l} F_{\m\n} g^{\k\m} g^{\l\n} \;\; ,\;\;
F_{\m\n} = \pa_\m A_\n - \pa_\n A_\m \;
\lab{F2-g} \\
L_D (\phi,g) = -\h g^{\m\n}\pa_\m \phi \pa_\n \phi - V(\phi) \; .
\lab{L-dilaton}
\er
In \rf{f-gravity+GG+D}--\rf{f-gravity} $R_{\m\n}(\G)$ indicates the Ricci curvature
in the first order (Palatini) formalism, \textsl{i.e.}, the space-time metric
$g_{\m\n}$ and the affine connection $\G^\m_{\n\l}$ are \textsl{a priori}
independent variables. Further notations indicate: 
$g \equiv \det\Vert g_{\m\n}\Vert$; the sign factor $\vareps = \pm 1$ in the square 
root term in \rf{GG-g} corresponds to ``magnetic'' or ``electric'' dominance; 
$f_0$ is a positive coupling constant. 

It is important to stress that:

(i) We will consider in what follows constant dilaton field, \textsl{i.e.},
ignoring the kinetic term in \rf{L-dilaton}. 

(ii) We will be particularly interested in the special cases $\L_0 = 0$
({\em no} bare cosmological constant term) and/or $e^2 \to \infty$ in \rf{GG-g},
\textsl{i.e.}, a nonlinear gauge field action \rf{GG-g} {\em without} a bare 
Maxwell term ($L(F^2(g))=-\frac{f_0}{2}\sqrt{\vareps F^2(g)}$).

As we will show below, both a standard kinetic Maxwell term for the gauge field 
with a variable $\phi$-dependent strength as well as a $\phi$-dependent cosmological
constant are {\em dynamically generated} as a combined
effect of the $R^2$-gravity term and the nonlinear ``square-root'' Maxwell term in
\rf{GG-g}.

The equations of motion resulting from the action \rf{f-gravity+GG+D} read:
\be
R_{\m\n}(\G) = \frac{1}{f^{\pr}_R}\llb \k^2 T_{\m\n} + 
\h f\bigl(R(g,\G)\bigr) g_{\m\n}\rrb 
\;\; ,\;\; f^{\pr}_R \equiv \frac{df(R)}{dR} = 1 + 2\a R(g,\G) \; ,
\lab{g-eqs}
\ee
\br
\nabla_\l \(\sqrt{-g} f^{\pr}_R g^{\m\n}\)  = 0 \; ,
\lab{gamma-eqs} \\
\pa_\n \Bigl(\sqrt{-g} \Bigl\lb 1/e^2 + \vareps \frac{f_0}{\sqrt{\vareps F^2(g)}}
\Bigr\rb F_{\k\l} g^{\m\k} g^{\n\l}\Bigr)=0 \; .
\lab{GG-eqs}
\er
About the dilaton, see Eq.\rf{V-extremum} below.
Here the total energy-momentum tensor is given by: 
\be
T_{\m\n} = \Bigl\lb L(F^2(g)) + L_D (\phi,g) - \frac{1}{\k^2}\L_0 \Bigr\rb g_{\m\n} 
+ \Bigl(1/e^2 + \frac{\vareps f_0}{\sqrt{\vareps F^2(g)}}\Bigr) 
F_{\m\k} F_{\n\l} g^{\k\l} + \pa_\m \phi \pa_\n \phi
\lab{T-total}
\ee
with $L(F^2(g))$  and $L_D (\phi)$ as in \rf{GG-g}--\rf{L-dilaton}.

Taking the trace of \rf{g-eqs} and using the explicit form of $f(R)$ in
\rf{f-gravity+GG+D} one gets for the Ricci scalar curvature:
\be
R(g,\G) = - \k^2 T (g) \quad , \quad T (g) = T_{\m\n} g^{\m\n} \; .
\lab{R-T-eq}
\ee
It can easily be shown 
(cf. \ct{olmo-etal})
that Eq.\rf{gamma-eqs} leads to the relation $\nabla_\l \( f^{\pr}_R g_{\m\n}\)=0$
and thus it implies transition to the ``physical'' Einstein-frame metrics 
$h_{\m\n}$ via conformal rescaling of the original metric $g_{\m\n}$:
\be
g_{\m\n} = \frac{1}{f^{\pr}_R} h_{\m\n} \quad ,\quad
\G^\m_{\n\l} = \h h^{\m\k} \(\pa_\n h_{\l\k} + \pa_\l h_{\n\k} - \pa_\k h_{\n\l}\)
\; .
\lab{einstein-frame}
\ee
Using \rf{einstein-frame} and taking into account \rf{R-T-eq} one can rewrite
gravity Eqs.\rf{g-eqs} within the Einstein frame as follows 
(from now on all space-time indices are raised/lowered by $h_{\m\n}$):
\be
R^\m_\n (h) = \k^2 \Bigl\lb \frac{1}{f^{\pr}_R} T^\m_\n (h) -
\frac{1}{4}\Bigl( 1 + \frac{1}{f^{\pr}_R}\Bigr)\, T(h) \d^\m_\n \Bigr\rb \; ,
\lab{einstein-like-eqs}
\ee
where:
\br
T_{\m\n} (h) = \Bigl\lb - \frac{1}{f^{\pr}_R}\bigl( V(\phi) + \L_0/\k^2\bigr) +
\(- \h f_0 \sqrt{\vareps F^2(h)} - X(\phi, h)\) \Bigr.
\nonu \\
\Bigl.
- f^{\pr}_R \frac{1}{4e^2} F^2(h)\Bigr\rb h_{\m\n}
+ \Bigl\lb \frac{f^{\pr}_R}{e^2} + \frac{\vareps f_0}{\sqrt{\vareps F^2(h)}}\Bigr\rb
F_{\m\k} F_{\n\l}h^{\k\l} + \pa_\m \phi \pa_\n \phi \; ,
\lab{T-h}
\er
\be
T(h) = T_{\m\n}(h) h^{\m\n} = - f_0 \sqrt{\vareps F^2(h)}
+ 2 X(\phi, h) - 4\frac{1}{f^{\pr}_R} \( V(\phi) + \L_0/\k^2\) \; ,
\lab{T-h-trace}
\ee
with short-hand notations:
\be
F^2(h) \equiv F_{\k\l} F_{\m\n} h^{\k\m} h^{\l\n}  \quad ,\quad
X(\phi, h) \equiv -\h h^{\m\n}\pa_\m \phi \pa_\n \phi \; ,
\lab{h-defs}
\ee
and
\br
f^{\pr}_R = \llb 1 + 2\a \k^2 T(h)\rrb^{-1} \phantom{aaaaaaaaaaaaaaaaaaaaaa}
\nonu \\
=
\Bigl\lb 1 + 8\a \(\k^2 V(\phi) + \L_0 \)\Bigr\rb \, \Bigl\lb 1 - 2\a \k^2
\( f_0 \sqrt{\vareps F^2(h)} + h^{\m\n}\pa_\m \phi \pa_\n \phi\)\Bigr\rb^{-1} \; .
\lab{f-prime}
\er
Accordingly, using \rf{f-prime} the nonlinear gauge field Eqs.\rf{GG-eqs} become:
\be
\pa_\n \(\sqrt{-h}\Bigl\lb \frac{1}{e_{\rm eff}^2 (\phi)} + 
\frac{\vareps f_{\rm eff}(\phi)}{\sqrt{\vareps F^2(h)}}\Bigr\rb 
F_{\k\l} h^{\m\k}h^{\n\l}\) = 0 \; ,
\lab{GG-eqs-h}
\ee
where we introduced the dynamical couplings:
\br
\frac{1}{e_{\rm eff}^2 (\phi)} = \frac{1}{e^2} - 
\frac{2\vareps\a\k^2 f_0^2}{1 + 8\a \(\k^2 V(\phi) + \L_0 \)} \; ,
\lab{e-eff} \\
f_{\rm eff} (\phi) = f_0 \frac{1+4\a\k^2 X(\phi, h)}{1 + 8\a \(\k^2 V(\phi) + \L_0 \)}
\; .
\lab{f-eff}
\er

Now, as an important step using the explicit expressions \rf{T-h}--\rf{f-prime}
we can cast Eqs.\rf{einstein-like-eqs} in the form of {\em standard} Einstein
equations:
\be
R^\m_\n (h) = \k^2 \({T_{\rm eff}}^\m_\n (h) - \h \d^\m_\n {T_{\rm eff}}^\l_\l (h)\)
\lab{einstein-h-eqs}
\ee
with energy-momentum tensor of the following form:
\be
{T_{\rm eff}}_{\m\n} (h) = h_{\m\n} L_{\rm eff} (h) - 2 \partder{L_{\rm eff}}{h^{\m\n}}
\lab{T-h-eff}
\ee
where (using notations \rf{e-eff}--\rf{f-eff}):
\br
L_{\rm eff} (h) = - \frac{1}{4 e_{\rm eff}^2 (\phi)} F^2(h) 
- \h f_{\rm eff} (\phi) \sqrt{\vareps F^2(h)} 
\nonu \\
+ \frac{X(\phi,h)\bigl(1+2\a\k^2 X(\phi,h)\bigr) - V(\phi)}{1+8\a\(\k^2 V(\phi)+\L_0\)}
\; .
\lab{L-eff-h}
\er
Thus, all equations of motion of the original $f(R)$-gravity/nonlinear-gauge-field 
system \rf{f-gravity+GG+D}--\rf{L-dilaton} with metric $g_{\m\n}$ can be 
equivalently derived from the following Einstein/nonlinear-gauge-field/dilaton action:
\be
S_{\rm eff} = \int d^4 x \sqrt{-h} \Bigl\lb \frac{R(h)}{2\k^2} 
+ L_{\rm eff} (h)\Bigr\rb \; ,
\lab{einstein-frame-action}
\ee
where $R(h)$ is the standard Ricci scalar of the metric $h_{\m\n}$ and 
$L_{\rm eff} (h)$ is as in \rf{L-eff-h}.

Let us particularly stress that in the absence of ordinary kinetic Maxwell term in
the original system ($e^2 \to \infty$ in \rf{GG-g}), such term is nevertheless
{\em dynamically generated} in the Einstein-frame action 
\rf{L-eff-h}--\rf{einstein-frame-action}:
\be
S_{\rm maxwell} =
-\h\a\k^2 f_0^2 \int d^4 x \sqrt{-h} \frac{F_{\k\l} F_{\m\n} h^{\k\m} h^{\l\n}}{
1+8\a\(\k^2 V(\phi)+\L_0\)} \; ,
\lab{dynamical-maxwell}
\ee
hence we will assume $\a >0$.
The Maxwell term generation occurs from the $R^2$ term in the original theory 
in the process of passing to the Einstein frame due to the on-shell relation \rf{R-T-eq}
where a {\em non-zero} trace of the 
energy-momentum tensor is produced by the nonlinear ``square-root'' gauge field term: 
$T(g) = - f_0 \sqrt{-F^2(g)} + (\phi-{\rm contribution})$.

\section{Non-Standard Black Hole Solutions}
\label{BH-nonstandard}

In what follows we will consider the case of constant dilaton $\phi$
extremizing the effective Lagrangian \rf{L-eff-h}, \textsl{i.e.}, we will ignore 
the kinetic $X(\phi, h)$-terms in \rf{L-eff-h}, and from now on we will concentrate
on the ``electric-dominant'' case $\vareps = -1$:
\br
L_{\rm eff} =
- \frac{1}{4 e_{\rm eff}^2 (\phi)} F^2(h) - \h f_{\rm eff} (\phi) \sqrt{-F^2(h)}
- V_{\rm eff}(\phi)
\lab{L-eff-0}
\er
Here the effective scalar potential and the effective couplings in \rf{L-eff-0}
are explicitly given by:
\br
V_{\rm eff}(\phi) = \frac{V(\phi) + \frac{\L_0}{\k^2}}{1+8\a\(\k^2 V(\phi)+\L_0\)}
\; ,
\lab{V-eff} \\
\frac{1}{e_{\rm eff}^2 (\phi)} = 
\frac{1}{e^2} + \frac{2\a\k^2 f_0^2}{1 + 8\a \(\k^2 V(\phi) + \L_0 \)} \; ,
\lab{e-eff-1} \\
f_{\rm eff} (\phi) = \frac{f_0}{1+8\a\(\k^2 V(\phi)+\L_0\)} \; .
\lab{f-eff-1}
\er
Since both $f_0$ and $f_{\rm eff} (\phi)$ (couplings of the ``square-root''
gauge field terms) must be positive (they determine the strength of the linear
confining part in the effective ``Cornell'' potential, cf. first ref.\ct{GG}
and \rf{lin-conf} below) we must have:
\be
1+8\a\(\k^2 V(\phi)+\L_0\) > 0 \; .
\lab{f-positive}
\ee

An important property of \rf{L-eff-0} is that the derivatives w.r.t. $\phi$ of 
the dynamical couplings \rf{e-eff-1}--\rf{f-eff-1}
are both extremized simultaneously with the extremization of the effective scalar
potential \rf{V-eff}:
\be
\partder{f_{\rm eff}}{\phi} = - 8\a\k^2 f_0 \partder{V_{\rm eff}}{\phi}
\;\; ,\;\; \partder{}{\phi}\Bigl(\frac{1}{e_{\rm eff}^2 (\phi)}\Bigr) =
-16\a^2\k^4f_0 \partder{V_{\rm eff}}{\phi} 
\quad \to \partder{L_{\rm eff}}{\phi} \sim \partder{V_{\rm eff}}{\phi} \; .
\lab{f-e-extremize}
\ee
Therefore at the extremum of $L_{\rm eff}$ \rf{L-eff-0} $\phi$ must satisfy:
\be
\partder{V_{\rm eff}}{\phi} = 
\frac{V^{\pr}(\phi)}{\llb 1+8\a\(\k^2 V(\phi)+\L_0\)\rrb^2} = 0 \; .
\lab{V-extremum}
\ee
There are two generic cases:

(a) Eq.\rf{V-extremum} is satisfied for some finite-value $\phi_0$ being an 
extremum of the original potential $V(\phi)$:
\be
V^{\pr}(\phi_0) = 0 \; .
\lab{phi-0}
\ee

(b) For polynomial or exponentially growing original $V(\phi)$, so that 
$V(\phi) \to \infty$ when $\phi \to \infty$, we have:
\be
\partder{V_{\rm eff}}{\phi} \to 0 \quad ,\quad 
V_{\rm eff} (\phi) \to \frac{1}{8\a\k^2} = {\rm const} \quad {\rm when} \;\;
\phi \to \infty \; ,
\lab{flat-region}
\ee
\textsl{i.e.}, for sufficiently large values of $\phi$ we find a ``flat region''
in the effective scalar potential \rf{V-eff}. Also, in this case we have
instead of \rf{e-eff-1}--\rf{f-eff-1}:
\be
f_{\rm eff} \to 0 \quad ,\quad e^2_{\rm eff} \to e^2
\lab{deconfine}
\ee
and \rf{L-eff-0} reduces to:
\be
L^{(0)}_{\rm eff} = -\frac{1}{4e^2} F^2(h) - \frac{1}{8\a\k^2} \; .
\lab{L-eff-h-0}
\ee

Now, the action \rf{einstein-frame-action} with the matter Lagrangian 
$L_{\rm eff} (h)$ as in \rf{L-eff-0} is of the same general form as the action of
the model describing ordinary Einstein gravity interacting with the nonlinear gauge
field theory containing square root of the Maxwell term, which was discussed
in refs.\ct{grav-cornell-hiding-hide-confine}, with the only difference being
the substitutions of the ordinary parameters $e^2,\, f_0,\,\L_0$ with the
effective $\phi$-dependent ones from \rf{V-eff}--\rf{f-eff-1}, where the
scalar field $\phi$ is constant. Therefore, we can implement the same steps
as in \ct{grav-cornell-hiding-hide-confine} to find static spherically
symmetric solutions of the system \rf{einstein-frame-action}, which is the
effective Einstein-frame form of the original 
$f(R)=R+\a R^2$-gravity/nonlinear-gauge-field theory \rf{f-gravity+GG+D}.

The static radial electric field $F_{0r}$ contains both Coulomb ($\sim \frac{1}{r^2}$)
as well as a {\em constant non-zero vacuum} piece (here and below $\phi={\rm const}$):
\be
|F_{0r}| = \Bigl(\frac{1}{e^2} + 
\frac{2\a\k^2 f_0^2}{1 + 8\a\(\k^2 V(\phi) + \L_0 \)}\Bigr)^{-\h}
\frac{f_0/\sqrt{2}}{1+8\a\(\k^2 V(\phi)+\L_0\)}
+\frac{|Q|}{\sqrt{4\pi}r^2}  \; .
\lab{h-cornell-sol}
\ee
The solution for the Einstein-frame metric $h_{\m\n}$ reads:
\be
ds_h^2 = - A(r) dt^2 + \frac{dr^2}{A(r)} + r^2 \bigl(d\th^2 + \sin^2 \th d\vp^2\bigr)
\; ,
\lab{h-spherical-static}
\ee
where:
\br
A(r) = 1 - \frac{\k^2 |Q|f_0}{\sqrt{8\pi} \llb 1+8\a\(\k^2 V(\phi)+\L_0\)\rrb}
- \frac{2m}{r} 
\nonu \\
+ \Bigl\lb\frac{1}{e^2}+\frac{2\a\k^2 f_0^2}{1+8\a\(\k^2 V(\phi)+\L_0\)}\Bigr\rb\,
\frac{\k^2 Q^2}{8\pi r^2} - \frac{\L_{\rm eff}(\phi)}{3} r^2 \; ,
\lab{h-metric-sol}
\er
with a total dynamical effective cosmological constant:
\be
\L_{\rm eff}(\phi) = \frac{\L_0 +\k^2 V(\phi)+\k^2 e^2 f^2_0/4}{
1+8\a\(\L_0 +\k^2 V(\phi)+\k^2 e^2 f^2_0/4 \)} \; .
\lab{h-CC-eff}
\ee
and with \textsl{a priori} free mass parameter $m$.

Therefore, in the case of ordinary extremum of the effective scalar potential
\rf{V-extremum}--\rf{phi-0} 
the properties of the solution
depend on the sign of the expression $\L_0 +\k^2 V(\phi_0)+\k^2 e^2 f^2_0/4$,
which determines the sign of $\L_{\rm eff}(\phi_0)$. We find:

(i) For positive/negative values of $\L_0 +\k^2 V(\phi_0)+\k^2 e^2 f^2_0/4$,
so that $\L_{\rm eff}(\phi_0) >0$ ($\L_{\rm eff}(\phi_0) <0$),
the solution \rf{h-cornell-sol}--\rf{h-CC-eff} describes 
Reissner-Nordstr{\"o}m-(anti-)de-Sitter-type black hole carrying an additional vacuum
radial electric field with magnitude:
\be
|E_{\rm vac}| = \Bigl(\frac{1}{e^2} + 
\frac{2\a\k^2 f_0^2}{1 + 8\a\(\k^2 V(\phi_0) + \L_0 \)}\Bigr)^{-\h}
\frac{f_0/\sqrt{2}}{1+8\a\(\k^2 V(\phi_0)+\L_0\)} \; .
\lab{vacuum-radial}
\ee

(ii) In the special case 
$\L_0 +\k^2 V(\phi_0)+\k^2 e^2 f^2_0/4 = 0$ when $\L_{\rm eff}(\phi_0)=0$, the
solution \rf{h-cornell-sol}--\rf{h-CC-eff} yields a
Reissner-Nordstr{\"o}m-type non-standard black hole which, apart from 
carrying the additional vacuum radial electric field \rf{vacuum-radial},
exhibits {\em non-flat} ``hedgehog''-type space-time asymptotics \ct{hedgehog}:
\be
A(r)\!\bgv_{r \to \infty}\! \simeq 
1 - \frac{\k^2 |Q|f_0}{\sqrt{8\pi}\llb 1+8\a\(\k^2 V(\phi_0)+\L_0\)\rrb} < 1 \; .
\lab{hedge-hog}
\ee

(iii) In the case of anti-de Sitter or hedgehog-type asymptotics the condition 
$\L_0 +\k^2 V(\phi_0)+\k^2 e^2 f^2_0/4 \leq 0$ together with \rf{f-positive} 
implies an upper bound for $\a$: $\a < \( 2\k^2 e^2 f^2_0\)^{-1}$.

(iv) Following the same steps as in Sect.4 of first 
ref.\ct{grav-cornell-hiding-hide-confine} we find a linear (w.r.t. $r$) 
confining piece in the effective potential of test charged particle dynamics
in the above black hole backgrounds:
\be
\frac{\sqrt{2}\cE |q_0|}{m_0^2} e_{\rm eff}(\phi_0) f_{\rm eff}(\phi_0)\, r \; ,
\lab{lin-conf}
\ee
where $\cE, m_0, q_0$ are energy, mass and charge of the test particle and
the effective gauge field couplings are as in \rf{e-eff-1}--\rf{f-eff-1}.

On the other hand, in the case of ``flat region'' for the effective scalar potential 
\rf{flat-region}--\rf{deconfine} 
the solution \rf{h-cornell-sol}--\rf{h-CC-eff} reduces to an {\em ordinary} 
Reissner-Nordstr{\"o}m-de-Sitter black hole:
\br
|F_{0r}| = \frac{|Q|}{\sqrt{4\pi}r^2} \quad, \quad
A(r) = 1 - \frac{2m}{r} + \frac{\k^2 Q^2}{8\pi r^2} - \frac{1}{24\a} r^2 \; ,
\lab{h-metric-sol-1}
\er
with an {\em induced} cosmological constant $\L_{\rm eff} = 1/8\a$, which is
{\em completely independent} of the bare cosmological constant $\L_0$.
Moreover, in the regime \rf{deconfine} the linear confining potential for
the test particles \rf{lin-conf} disappears.

\section{Generalized Levi-Civita-Bertotti-Robinson Solutions}
\label{tubelike}

Following the same steps as in second and third 
refs.\ct{grav-cornell-hiding-hide-confine} we can find
explicit static solutions of generalized  Levi-Civita-Bertotti-Robinson (LCBR) 
type \ct{LC-BR} of the system \rf{einstein-frame-action}--\rf{L-eff-0}. 
For definiteness we will concentrate on the case of ``electric dominance'' and for 
simplicity we will use units with Newton constant $G_N=1$, \textsl{i.e.},
$\k^2 = 8\pi$, and $e^2=1$. These generalized LCBR-type space-times are ``tube-like''
solutions with space-time geometry of the form $\cM_2 \times S^2$ where $\cM_2$ 
is some two-dimensional manifold with coordinates $(t,\eta)$:
\be
ds_h^2 = - A(\eta) dt^2 + \frac{d\eta^2}{A(\eta)} 
+ r_0^2 \bigl(d\th^2 + \sin^2 \th d\phi^2\bigr) \;\; ,\;\;  
-\infty < \eta <\infty \;\; ,\;\; r_0 = \mathrm{const} \; .
\lab{gen-BR-metric}
\ee
and with a constant radial static electric field:
\be
F_{0\eta} = c_F = \mathrm{arbitrary ~const} \; .
\lab{const-electr}
\ee

The Einstein equations corresponding to \rf{einstein-frame-action}--\rf{L-eff-0}
yield the following relation for $r_0$:
\be
\frac{1}{r_0^2} = \frac{4\pi}{1+8\a\L(\phi_0)}\Bigl\lb
\Bigl(1+8\a\(\L(\phi_0)+2\pi f_0^2\)\Bigr) c_F^2 +\frac{1}{4\pi}\L(\phi_0)\Bigr\rb
\lab{r0-eq}
\ee
with the short-hand notation 
$\L(\phi_0)\equiv 8\pi V(\phi_0) + \L_0 \; ,$
and the following simple differential equation for the metric coefficient
$A(\eta)$ in \rf{gen-BR-metric}:
\br
\pa_\eta^2 A(\eta) = \frac{8\pi}{1+8\a\L(\phi_0)} K(c_F) \; ,
\lab{A-diff-eq} \\
K(c_F) \equiv \Bigl(1+8\a\(\L(\phi_0)+2\pi f_0^2\)\Bigr) c_F^2 - \sqrt{2}f_0 |c_F| - 
\frac{1}{4\pi}\L(\phi_0) \; .
\lab{K-def}
\er
In the case of ``flat region'' of the effective scalar potential \rf{flat-region}
$\L(\phi_0) \to \infty$, so that Eqs.\rf{r0-eq}--\rf{A-diff-eq} simplify:
\be
\frac{1}{r_0^2} = 4\pi c_F^2 + \frac{1}{8\a} \quad ,\quad
\pa_\eta^2 A(\eta) = 8\pi c_F^2 - \frac{1}{4\a} \; .
\lab{flat-region-eqs}
\ee

As in last ref.\ct{grav-cornell-hiding-hide-confine}, there are three distinct
types of generalized LCBR solutions depending on the sign of the factor $K(c_F)$
in \rf{A-diff-eq}--\rf{K-def}.

\vspace{.1in}
(A)  $AdS_2 \times S^2$ with strong constant vacuum electric field
$F_{0\eta}=c_F $, where $AdS_2$ is two-dimensional anti-de Sitter 
space with:
\be
A(\eta) = 4\pi K(c_F) \eta^2 \quad ,\quad K(c_F) >0 \; ,
\lab{AdS2}
\ee
in the metric \rf{gen-BR-metric}, $\eta$ being the Poincare patch
space-like coordinate. The magnitude $|c_F|$ of the vacuum electric field
must satisfy the inequalities:
\be
|c_F| > \frac{f_0}{\sqrt{2}\Bigl\lb 1+8\a\(\L(\phi_0)+2\pi f_0^2\)\Bigr\rb}
\llb 1 + \sqrt{1 + \frac{\L(\phi_0)}{2\pi f_0^2}
\Bigl\lb 1+8\a\(\L(\phi_0)+2\pi f_0^2\)\Bigr\rb}\rrb
\lab{AdS2-1}
\ee
for $\L(\phi_0) > {\rm max} \lcurl -2\pi f_0^2,\, - \frac{1}{8\a} \rcurl$;
\be
c^2_F > \frac{|\L(\phi_0)|}{4\pi \Bigl\lb 1+8\a\(\L(\phi_0)+2\pi f_0^2\)\Bigr\rb}
\quad ,\;\; {\rm for} \;\; - \frac{1}{8\a} < \L(\phi_0) < - 2\pi f_0^2 \; .
\lab{AdS2-2}
\ee
In the ``flat region'' case \rf{flat-region-eqs} $|c_F| > (32\pi \a)^{-\h}$.

\vspace{.1in}
(B) $Rind_2 \times S^2$ with constant vacuum electric field $F_{0\eta}= c_F$
when the factor $K(c_F)=0$. Here $Rind_2$ is the flat two-dimensional 
Rindler space with:
\be
A(\eta) = \eta \;\; \mathrm{for}\; 0 < \eta < \infty \quad \mathrm{or} \quad
A(\eta) = - \eta \;\; \mathrm{for}\; -\infty <\eta < 0 
\lab{Rindler2}
\ee
in the metric \rf{gen-BR-metric} and:
\be
|c_F| = \frac{f_0}{\sqrt{2}\Bigl\lb 1+8\a\(\L(\phi_0)+2\pi f_0^2\)\Bigr\rb}
\llb 1 + \sqrt{1 + \frac{\L(\phi_0)}{2\pi f_0^2}
\Bigl\lb 1+8\a\(\L(\phi_0)+2\pi f_0^2\)\Bigr\rb}\rrb
\lab{Rindler2-1}
\ee
for $\L(\phi_0) > {\rm max} \lcurl -2\pi f_0^2,\, - \frac{1}{8\a} \rcurl$.
In the ``flat region'' case \rf{flat-region-eqs} $|c_F| = (32\pi \a)^{-\h}$.

\vspace{.1in}
(C)  $dS_2 \times S^2$ with weak constant vacuum electric field
$F_{0\eta}= c_F$, where $dS_2$ is two-dimensional de Sitter space with:
\be
A(\eta) = 1 - 4\pi |K(c_F)|\,\eta^2 \quad ,\quad K(c_F) <0 \; ,
\lab{dS2}
\ee
in the metric \rf{gen-BR-metric}. The magnitude $|c_F|$ of the vacuum electric field
must satisfy:
\be
|c_F| < \frac{f_0}{\sqrt{2}\Bigl\lb 1+8\a\(\L(\phi_0)+2\pi f_0^2\)\Bigr\rb}
\llb 1 + \sqrt{1 + \frac{\L(\phi_0)}{2\pi f_0^2}
\Bigl\lb 1+8\a\(\L(\phi_0)+2\pi f_0^2\)\Bigr\rb}\rrb \; ,
\lab{dS2-1}
\ee
where $\L(\phi_0) > {\rm max} \lcurl -2\pi f_0^2,\, - \frac{1}{8\a} \rcurl$.
In the ``flat region'' case \rf{flat-region-eqs} $|c_F| < (32\pi \a)^{-\h}$.

\section{Discussion}
\label{discuss}

In the present note we have considered $f(R)=R + \a R^2$-gravity within the
first-order (Palatini) formalism coupled to dilaton and a special kind of
nonlinear gauge field system containing a square-root of the standard
Maxwell term, which is known to produce a QCD-like confinement. 
We have derived the explicit form of the dynamically equivalent ``physical'' 
Einstein-frame effective theory displaying the following significant properties:

(i) The effective gauge field couplings as well as the effective cosmological
constant become functions of the constant dilaton in such a way that even in the
event of absence of kinetic Maxwell term for the gauge field and/or absence of bare
cosmological constant in the original theory, the latter are nevertheless
dynamically generated in the Einstein-frame effective theory.

(ii) There are two interesting regimes for the constant dilaton $\phi$ : 
(a) either as a minimum of the bare scalar potential at some finite value 
$\phi_0$, which coincides with the minimum $\phi_0$ of the effective scalar
potential in the Einstein-frame theory, or (b) $\phi$ belongs to a ``flat region'' 
of the effective scalar potential, which
corresponds to a fast growing at infinity bare scalar potential. In the
first case the effective coupling of the confinement-producing
``square-root'' gauge field term remains finite, whereas in the second case
it vanishes and so does the confining feature. This picture resembles the
``MIT bag'' structure \ct{MIT-bag} where inside the ``bag'' a regular
gauge-field dynamics holds, while outside the ``bag'' the gauge fields are
confined. Moreover, the effective dilaton potential with a ``flat region''
can be used for inflation.

(iii) The effective coupling constants in the Einstein-frame theory satisfy
the ``least coupling principle'' of Damour-Polyakov \ct{damour-polyakov},
namely, any extremal point $\phi_0$ of the scalar effective potential (as function
of the dilaton) is simultaneously an extremal point of the effective gauge
couplings. This property is crucial for the consistency of the solutions
here obtained.

(iv) We derived new solutions describing non-standard black holes and
Levi-Civitta-Bertotti-Robinson-type ``tube-like'' space-times, generalizing
those found in \ct{grav-cornell-hiding-hide-confine}, in that now the
pertinent constant vacuum electric fields and the non-flat ``hedgehog''
space-time asymptotics depend on the dilaton value $\phi_0$.

Let us also point out that an $R^2$-gravity theory coupled to nonlinear
gauge field system with Maxwell and ``square-root'' terms and a dilaton was
earlier studied in the context of the so called two-measure gravity models 
in ref.\ct{TMT-R2}. The results there about the explicit form of the
Einstein-frame effective theory resembles those obtained in the present note.

\section*{Acknowledgments}
E.G. is thankful to  D. Rubiera-Garcia and G. Olmo for stimulating discussions.
E.N. and S.P. are supported by Bulgarian NSF grant \textsl{DO 02-257}.
Also, all of us acknowledge support of our collaboration through the exchange
agreement between the Ben-Gurion University 
and the Bulgarian Academy of Sciences.


\end{document}